\documentclass[widereview]{acmsiggraph}        

\usepackage[scaled=.92]{helvet}
\usepackage{times}
\usepackage{graphicx}
\usepackage{amsmath}
\usepackage{bm}
\usepackage{parskip}
\usepackage[labelfont=bf,textfont=it]{caption}
\usepackage{color}
\usepackage{subfigure}

\title{CanvoX: High-resolution VR Painting in Large Volumetric Canvas}

\author{Yeojin Kim\thanks{e-mail:yeojinkim@ewhain.net}\\Ewha Womans University %
\and Byungmoon Kim\thanks{e-mail:bmkim@adobe.com}\\Adobe Research %
\and Jiyang Kim\thanks{e-mail:soarmin11@ewhain.net}\\Ewha Womans University %
\and Young J. Kim\thanks{e-mail:kimy@ewha.ac.kr}\\ Ewha Womans University}

\onlineid{0202}

\begin{document}

\teaser{
    \includegraphics[height=4cm]{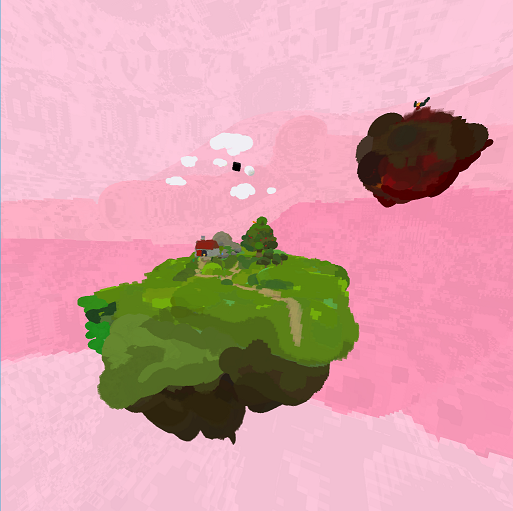} \hspace{0.1cm}
    \includegraphics[height=4cm]{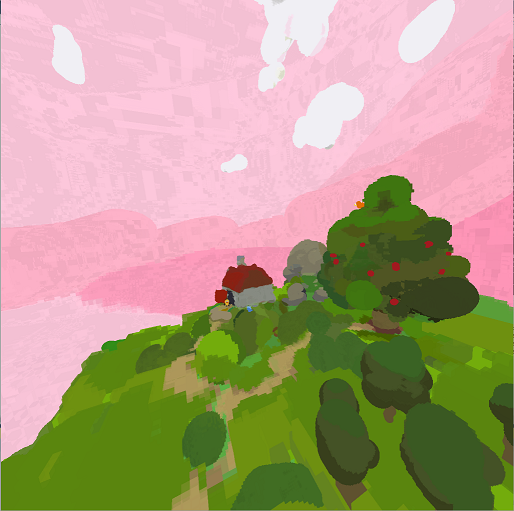} \hspace{0.1cm}
    \includegraphics[height=4cm]{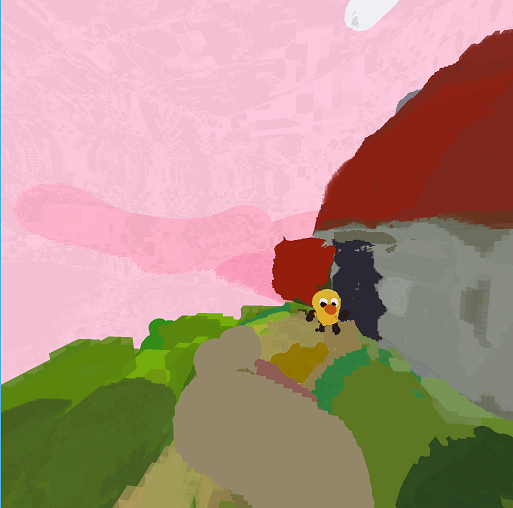} \hspace{0.1cm}
    \includegraphics[height=4cm]{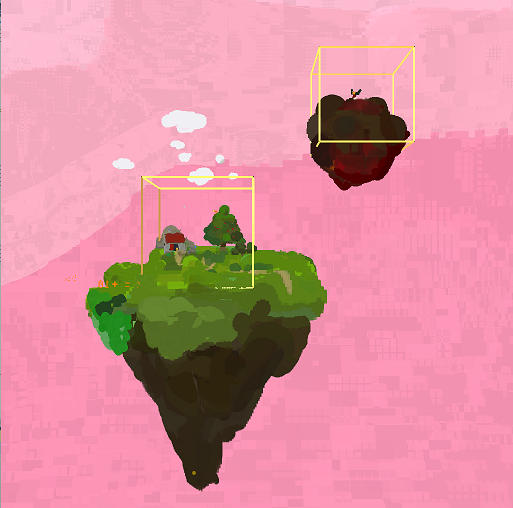}%
  \caption{Virtual reality painting in very large canvas (40 km$^3$) and with high details (down to 0.3 mm$^3$). Background is a large volumetric structure painted by a very large brush (about one kilometers), while foreground objects are painted with a very fine (a few millimeters) brush. To help audiences not to lost in the painting, and to allow artist to focus, foreground area, called {\it rooms} can be defined, shown as yellow boxes in the far right image. Audiences starts from a room, navigate (translate and scale) in the canvas, and switch to another room any time.}
  \label{fig:teaser}
}

\maketitle


\begin{abstract}
With virtual reality, digital painting on 2D canvases is now being extended to 3D spaces. Tilt Brush and Oculus Quill are widely accepted among artists as tools that pave the way to a new form of art - 3D emmersive painting. Current 3D painting systems are only a start, emitting textured triangular geometries. In this paper, we advance this new art of 3D painting to {\em 3D volumetric painting} that enables an artist to draw a huge scene with full control of spatial color fields. Inspired by the fact that 2D paintings often use vast space to paint background and small but detailed space for foreground, we claim that supporting a large canvas in varying detail is essential for 3D painting. In order to help artists focus and audiences to navigate the large canvas space, we provide small artist-defined areas, called {\it rooms}, that serve as beacons for artist-suggested scales, spaces, locations for intended appreciation view of the painting. Artists and audiences can easily transport themselves between different rooms. Technically, our canvas is represented as an array of deep octrees of depth 24 or higher, built on CPU for volume painting and on GPU for volume rendering using accurate ray casting.  
In CPU side, we design an efficient iterative algorithm to refine or coarsen octree, as a result of volumetric painting strokes, at highly interactive rates, and update the corresponding GPU textures. Then we use GPU-based ray casting algorithms to render the volumetric painting result. We explore precision issues stemming from ray-casting the octree of high depth, and provide a new analysis and verification. From our experimental results as well as the positive feedback from the participating artists, we strongly believe that our new 3D volume painting system can open up a new possibility for VR-driven digital art medium to professional artists as well as to novice users.
\end{abstract}

\section{Introduction}\label{sec:intro}

Since prehistoric days human has been painting on 2D canvas. The oldest painting can be traced back to around 40,000 years ago, made by a homo sapience in El Castillo, and around the same time, a Neanderthal's rock engraving in Gorham's cave was made. Throughout human history, pioneering artists have been continuously searching for new art forms, in particular, new forms of paintings. Certain forms gained popularity in time - an example would be Van Gogh style paintings. With virtual reality (VR), digital painting on 2D surfaces is being extended into 3D volumes. VR-based 3D painting applications such as Tilt Brush and Quill have recently surged and are now widely accepted by artists to be positioned as a new art form. 3D painting is distinguished from real world sculpting as there are no limits posed by real world material or gravitational constraints, while VR systems provide realistic or non-realistic emmersive view of the new art form. We believe these new VR-based 3D paintings will thrive, similarly to 2D paintings started from the prehistoric cave-wall era.

However, current 3D painting is in an early stage, and has much room for improvement. State-of-the-art systems such as Tilt brush and Quill output only mesh geometry as a result of artist's stroking, and relies on a rasterization pipeline to render the geometry. Such system is cannot easily depict thick volumetric shapes and tends towards being a collection of thin ribbon likes structures.
Color mixing of multiple brush-strokes is hard to implement and not supported by current systems; e.g. Fig.~\ref{fig:StrokeMixComparison}. Since each stroke needs to add mesh-geometry, the number of applicable strokes is bounded by the number of strokes a GPU can render. When an artist paints over and over to correct or mix colors, there will be no color mixing effect and this only increases the stroke count which steadily degrades system performance. Therefore a common painting task of repetitive stroking is not yet available in existing 3D painting systems. Finally, semi-transparent strokes are hard to render as a large number of long and highly curved strokes may complicate the depth-ordering. 


\paragraph{Main Results} In this paper, we explore a new route for VR painting in 3D space, {\em volumetric painting}. In contrast to volume-less, open-surface strokes, used by the state-of-the-art 3D painting systems, our volumetric strokes applied to a 3D grid is amenable to depicting both solid and non-solid shapes.  Moreover, artists can repeatedly apply strokes to the same area until satisfied with the mixed color result and its detail; without having appreciable degradation in performance. Transparency is also naturally handled.
In a 3D volumetric canvas, unlike 2D planar canvas, distant objects in the background can be painted relatively large in size compared to foreground objects  - for example distant mountains. To support large canvases, we use an array of deep octrees of high depth (e.g. level 24 or higher). We then define {\em rooms} to allow detailed paintings for foreground objects. The remainder of the scene (outside the defined rooms) serves as background.  We can limit details that are not observable from any room. In this way, we can maintain very large canvas, e.g. a virtual canvas of up to 40 Km$^3$ painting space with very fine details of 0.3mm$^3$ painting of tiny voxels inside a room (see Fig.~\ref{fig:teaser}). Artists can also define multiple rooms anywhere inside the 3D canvas and add more details when observable from the rooms. In summary, our main contributions to VR painting are:
\begin{itemize}
    \item Volumetric painting using large canvas with high detail,
    \item Large canvas painting with varying resolution based on distance or visibility from rooms,
    \item Rooms for multiple room-scale appreciations/experience.
\end{itemize}

To realize this idea on commodity computing and VR platforms, we propose a new painting system capable of incremental updates to the underlying GPU-based octree data structures along with a novel stereo volume rendering techniques using accurate ray casting. In detail, our painting system addresses the following technical challenges:
\begin{itemize}
    \item A hybrid octree data structure that utilizes both CPU and GPU architectures and maintains efficient synchronization between them,
    \item Incremental budget-limited octree refinement at highly-interactive rates for painting and rendering,
    \item Precision analysis using cell-local coordinates to allow accurate ray casting for deep octree evaluation.
\end{itemize}

The rest of this paper is organized as follows. Followed by brief literature survey in section \ref{sec:prev}, we discuss our observations on 3D painting in comparison to 2D painting and 3D modeling in section \ref{sect:3dpainting}.  In section 4, we describe our adaptive grid representation and explain our painting modeling and VR interfaces in section 5. In section 6, we provide an accurate ray-casting algorithm for rendering adaptive grids and discuss experimental results in section 7. We conclude our paper and discuss future work in section 8.

\section{Related Work}\label{sec:prev}

We briefly survey the prior works relevant to our new 3D painting system including GPU-based octree representations and ray-casting on them.

\subsection{3D Painting System}
The concept of 3D painting system was introduced for more than two decades ago. In early 90s, \cite{hanrahan1990:3DPainting} showed that artists can paint directly on 3D models in real-time, but their work is incapable of representing transparency and fine detailed painting. From ~\cite{Daily1995:3DPainting}, 3D painting systems started storing data in a 2D texture map to support painting details. However, 2D parametrization leads to artifacts such as distortion or seams. To avoid these problems, ~\cite{DeBry2002,Benson2002} suggested an octree representation based on 2D texture map which contains only the surface of 3D models. With the increasing performance of graphics hardware, it is possible to accelerate these algorithms by porting them from the CPU to GPU~\cite{Lefebvre2005} or to implement octree-like data structures entirely on GPU~\cite{GLIFT:lefohn2006}. More recently, ~\cite{Overcoat:Schmid:2011} proposed implicit canvas that uses parametric strokes to be projected onto the underlying 3D models. Note that all of these 3D painting systems focus only on painting the 2D surfaces of models, not extended to volumetric painting, mainly due to high memory consumption.

With the advent of modern VR systems, it become very natural to extend 2D painting to 3D space; beginning with ~\cite{Keefe:2001}'s work, VR painting systems such as Tilt Brush\footnote{http://www.tiltbrush.com} and Quill\footnote{http://storystudio.oculus.com} lead a new trend for 3D painting and modeling. All these systems render quad strips as a result of painting strokes, thus have a difficulty in representing volumetric objects. 
Furthermore, voxel-based modeling or painting such as Oculus Medium\footnote{http://www.oculus.com/medium/} or High Fidelity\footnote{http://highfidelity.io} were introduced. However, due to the space and time complexity of uniform grid, they support only limited canvas spaces and also limited detail.  

\subsection{GPU-based Octree}
The idea of a GPU-based octree representation starts from ~\cite{Lefebvre2005} with pointer pools, assuming that the underlying model is static. This work supports top-down traversal on an octree.
Other researchers~\cite{gobbetti2008single,GIGAVOXEL:Crassin,SVO:Laine:2010,museth2013vdb} extended the octree application to massive volumetric data sets. ~\cite{gobbetti2008single,GIGAVOXEL:Crassin} manages the octree both on CPU and GPU asynchronously in a view-dependent manner. Glift~\cite{GLIFT:lefohn2006} explored an adaptive resolution tile packed to seamlessly cover a domain entirely on the GPU with constant-time access.

\subsection{Ray Casting on Adaptive Grids}
Existing ray-casting algorithms for adaptive grids often utilize acceleration techniques, such as empty space skipping~\cite{kruger2003acceleration}, by using multiple rendering passes.
In order to reduce neighbor finding cost, octree neighbor linking were proposed ~\cite{samet1989octree,macdonald1990heuristics}. Also the ROPE algorithm ~\cite{ROPE:havran1998ray} was developed to accelerate k-d tree neighbor finding. These precomputed neighbors have been continuously used. ~\cite{gobbetti2008single} suggests stackless ray-casting after updating all six neighbors from the CPU to the GPU in view-dependent manner. \cite{GLIFT:lefohn2006,GIGAVOXEL:Crassin} used similarly precomputed neighbors to accelerate neighbor finding for samples stored at octree corners. In this paper, we make a small step forward by using only three neighbors, computed in the GPUs from the primal octree represented by only two indices: parent and the first child. 

To our best knowledge, octrees as deep as 24 were not used for ray casting. The deepest trees we found was $4096^3$ used in ~\cite{museth2013vdb}, which is equivalent to the octree depth of 12 (our canvas is equivalent to $(4\times2^{24})^3$), and the ray angle drift error has not been identified as a challenge. We propose a solution and thoroughly analyze the numerical precision associated with ray drift during ray traversal using a very deep tree.

\section{Observations on 3D Painting} \label{sect:3dpainting}
\subsection{Transitioning From 2D To 3D Painting}

The space representable by 2D paintings is not limited - it is vast while detailed. Imagine an artist wanting to paint a bacteria sitting at the tip of a needle with the Orion constellation in the background. In a similar vein, 3D painting should not have spatial or scale limits.
However, unlikely in 2D paintings that typically have details only for near objects with a single view/appreciation point and no 3D-depth, in 3D painting, highly detailed objects can exist at multiple places in the canvas by placing the audience appreciation spots anywhere in 3D. Even though artists cannot paint details everywhere, they should be able to paint details in a few places. In fact, navigating the canvas while zooming in-out to a large space is already supported by state-of-the-art systems such as Tilt Brush and Quill. Artists are already exploring high powered zooming options demonstrated by the recent painting {\it worlds-in-worlds}\footnote{http://youtu.be/EzsG1uqfDTQ}, although audience may not be able to locate the details as no hint is provided for a point of interest and scale. 

\subsection{Comparing 3D Painting and Modeling/Rendering}
Painting large canvases in high detail appears to be one of the aspects in 3D paintings different from 3D modeling, which typically focuses on modeling an object at a given scale precisely, rather than zooming in or out by a large amount. In conventional 3D graphics pipelines, a scene is constructed by a team of artists modeling objects, assigning textures and materials, placing lights, and eventually compositing them; in 3D painting, an individual artist alone needs to paint the entire scene. An artist should be able to do so in an unconstrained manner, i.e., artists should have a direct and intuitive control of coloring at any location. If they want to express their artistic inspiration to depict a specific way of lighting, they should be able to paint lighting effects completely the way they want. Thus, automatic lighting or shading is not a must - although it will be useful.

Interestingly, in VR-based paintings, artists enjoy expressiveness, owing to the large virtual space; e.g. the artists  tend to make large, initial strokes. They often prefer VR hand controllers to haptic styluses with higher precision but with much smaller configuration space. Preference of expressiveness to precision is perhaps another aspect that differentiates 3D painting from 3D modeling.

\section{Adaptive and Large Volumetric Grid}
It appears to be rather trivial for the surfaced-based 3D painting systems such as Tilt Brush to support large and highly zoomable canvases. However, this requirement would be a major challenge in volume painting; existing uniform voxel editing engines, e.g. such as Minecraft\footnote{http://minecraft.net} cannot meet this requirement. To allow large and detailed canvases, we use an array of {\em deep octrees}. Each array element is an octree root that can be refined 24 times or higher. While the root array helps reducing tree depth and enables immediate parallelization, all roots that overlap with a brush stroke must be visited.
Therefore, we resort to a relatively coarse, $4^3$ array of octree roots, each of which can be refined to a maximum depth; effectively, this is equal to an octree of depth 26, hence a deep octree. Using this octree array, we obtain canvas in $40Km^3$ w.r.t. the size of a room-scale VR environment with maximum details of $0.3mm^3$ voxels. 

\subsection{Hybrid Representation}
We use both CPU and GPU to represent deep octrees for the following reasons. As CPU alone is not efficient enough to render our large volumetric canvas in stereo for both eyes, each at resolution 1680 $\times$ 1512 and 90 FPS, using the GPU becomes an obvious choice. CPU is primarily used for maintaining and adjusting the octree, something that is non-trivial to implement on the GPU, especially using OpenGL. The octrees on the CPU are synchronized with those in GPU in a deferred fashion to maintain high interactive rates.
In addition we found painting strokes on the CPU more flexible, since we have the ability to perform blend modes not supported natively by GPU hardware.

\subsubsection{CPU Memory Layout}
Since we are authoring volumetric fields defined everywhere and also for the sake of simplicity, when a cell is refined, we always create eight children. Similarly to ~\cite{gobbetti2008single}, we do not use pointers, but instead use the index $I$ that uniquely identifies each cell. Our octree is made up of multiple linear memory pools indexed by $I$. Painting properties such as color, density, and temporary variables are stored as separate field pools. Field pools can be even added dynamically if needed. We group properties that are likely to be accessed together and then store them in a single memory pool. For example, we have a pool that stores the color as well as alpha of a painted voxel packed per each cell. To allow dynamic refinement and coarsening, we implement linked-list based memory management. The size of an allocation unit is fixed as we allocate or free eight cells. When a unit is freed, we return to the beginning of the pool so that the pool is populated from the beginning. The pool begins with a uniform root array that cannot be freed. We have another separate flag pool for cell depth and other bit field flags. 

\subsubsection*{Tree Graph ${\cal G}_p$ and Neighbor Graph ${\cal G}_3$} \label{sec:treetopologies}
Our tree is defined only by parents and children pools that store the indices of parents and the first children. Note that remaining seven children's indices are consecutively numbered and hence need not be stored. Let this primal tree graph be ${\cal G}_p$. Since we do not use a dual tree ~\cite{GLIFT:lefohn2006,SVO:Laine:2010}, ${\cal G}_p$ is represented only by two 32-bit indices: parent and the first child, at each cell. 

\begin{figure}
    \centering
    \includegraphics[width=0.55\linewidth]{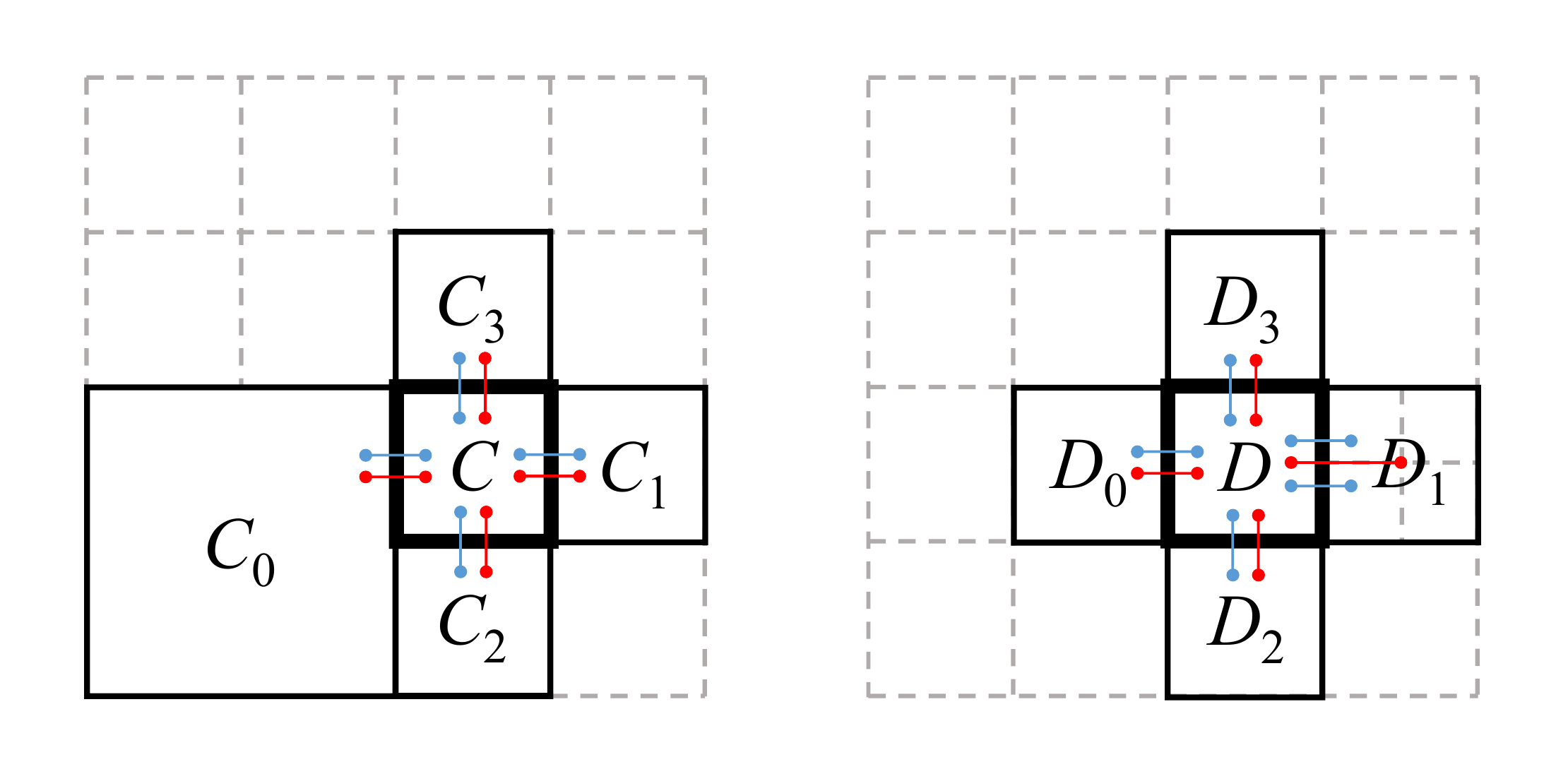}
    \caption{Neighbors of C (left), and D (right) in quadtree. In this figure, $C,C_1,C_2$ share the same parent, and hence computing $C_1$ and $C_2$ from $C$ is easy. On the other hand, primal tree distances between $C$ and $C_0,C_3$, and between $D$ and $D_1,D_3$ can be very large, and hence we precompute them, and store them in a separate texture.}
    \label{fig:neighbor}
\end{figure}

Note that ray casting using only ${\cal G}_p$ may have poor performance as the depth of tree increases; e.g. if the maximum depth is 24, the distance in ${\cal G}_p$ to a neighbor can be as long as 48. Therefore, an immediate neighbor topology such as neighbor linking for octrees \cite{samet1989octree,macdonald1990heuristics}, or ropes for KD trees ~\cite{ROPE:havran1998ray}, is useful to accelerate ray traversal. ROPE connects cells sharing a face, resulting in the blue connections from cells $C$ and $D$ in Fig. \ref{fig:neighbor}. $D$ has five neighbors. In 2-to-1 balanced octrees \cite{kim2015interpolation}, the number of neighbors can vary from 6 to 24 since one or four neighbors exist per each of the six cell faces. In our work, we reduce the number of neighbors down to three.

First, consider the same depth cell (e.g. $D_1$ in Fig.~\ref{fig:neighbor}) or smaller-depth neighboring cell ($C_0$). 
Since each face has only one such neighbor, every cell has six neighbors. We store only three of these six neighbors. Since eight children have consecutive indices, three of such neighbors that share the same parent (e.g. $C_1,C_2$ of $C$ and $D_0, D_2$ of $D$) and can be obtained immediately from its associated cell index (e.g., $C_1=C+1$). The other three neighbors may have different parents (e.g. $C_0,C_3,D_1,D_3$), and computing such neighbors requires tree traversals, and hence they are precomputed. We refer to this neighbor structures as the 3-neighbors topology, generated from the graph denoted by ${\cal G}_3$ that is a collection of the three neighbors for each cell. 

Using the union of ${\cal G}_3$ and ${\cal G}_p$, we can quickly discover all the six to 24 neighbors, since in ${\cal G}_p\cup{\cal G}_3$, all the two cells sharing a face have distance 2 in ${\cal G}_p$, or 1 or 2 in ${\cal G}_3$. Therefore, finding neighbors has a low cost. For example, the leaves of $D_1$ that contact with $D$ can be found easily as ${\cal C}(D_1)$ and ${\cal C}(D_1)$+2, where ${\cal C}(\cdot)$ denotes the first child. In case a ray proceeds to $D_1$, we can quickly identify ${\cal C}(D_1)$ or ${\cal C}(D_1)+2$ using the ray parameters. Since ${\cal G}_3$ is represented by three indices, we store ${\cal G}_3$ as a separate neighbor pool. 

\subsubsection{GPU Memory Layout}
Mapping the octree pools in CPU to GPU textures is straightforward. Since GPU textures has a resolution-limit in each dimension, we cannot use a 1D texture. We must use 2D or 3D textures and map a linear index $I$ to two or three indices. Since modern GPUs support up to $16$ thousand texels per dimension, 2D textures will support up to $256M$ cells. We pack ${\cal G}_p$ (parent, child) and depth into a texture, ${\cal G}_3$ (3-neighbors) into another texture, and have RGBA color stored in the other texture. Note that we do not use ${\cal G}_3$ on the CPU. ${\cal G}_3$ texture is rendered from the ${\cal G}_p$ texture upon updates.

\section{Volume Painting and Interfaces}\label{sect:painting}

\subsection{Color Mix, Pick-Up, and Brush Stamps}
In digital painting, pigment deposition from brush and mixing with canvas color itself is a separate topic and is beyond the scope of this paper. Fortunately, color mixing developed in 2D digital painting applications is directly applicable to 3D painting. In this paper, we only implement a very simple additive blending mode. Given brush color ($B_r,B_g,B_b,B_\alpha$) and canvas color ($C_r,C_g,C_b,C_\alpha$), the output color will be $(B_r,B_g,B_b)m + (1-m)*(C_r,C_g,C_b)), m = b_\alpha / (b_\alpha+C_\alpha)$, and the output opacity is $b_\alpha+c_\alpha$. Note that there will be many other alternatives such as the popular per-stroke maximum-opacity based model \cite{Photoshop}, physically-based pigment mixing using Kubelka-Munk mode\ \cite{Baxter04}, RYB mixing \cite{WETBRUSH:Chen:2015}, or advanced RGB-space color mixing  \cite{Lu:2014:RPC}.

Similarly, 2D brush stamping methods (another orthogonal topic) can be directly applied to 3D painting. We currently support multiple stamp shapes: sphere, cylinder, box, cone, and procedural Perlin noise. For spherical stamping, we support sweeping, resulting in tapered capsules. We place these tapered capsules to connect the two consecutive samples of a stroke path. 

\begin{figure}[htb]
\centering
    \subfigure[Our system]{\includegraphics[height=4.3cm]{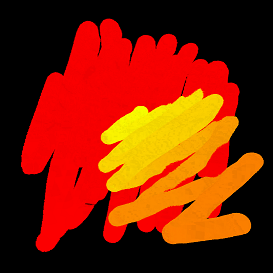}\label{fig:MixVolume}}
    \subfigure[Tilt Brush]{\includegraphics[height=4.3cm]{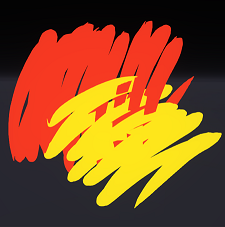}\label{fig:MixTiltBrush}}
\caption{Comparison in terms of color-mix of red and yellow brush strokes.}
\label{fig:StrokeMixComparison}
\end{figure}

Another common 2D painting practice is color pick-up. We demonstrate this with a simple pick-up implementation. At each sampling point, a brush can pick up color from canvas and blend it with the current brush color. Similarly to most 2D painting system \cite{Photoshop}, when a stroke is complete, we restore the brush color to the original brush color. As show in Fig. \ref{fig:StrokeMixComparison} (a), color pick-up automatically generates spatially-varying colors, and this is a popular way that artists generate color variation for painting. In contrast, surface-based painting systems like Tilt Brush does not have color mixing between strokes as shown in Fig. \ref{fig:StrokeMixComparison} (b).

\subsection{Adaptive Brush Size}
In 2D painting, artists often want to apply brush-strokes fast and react to immediate feedback. This is simply continued in 3D painting. For 2D painting, modern CPUs can handle reasonably large brush sizes of up to a few hundred pixels at interactive rates, but for a larger brush, interactivity starts to diminish. This could be quite restrictive as well as frustrating to artists; when an artist paints on a large 2D canvas, the artist has to suffer the delay in a larger brush or should use a smaller brush to fill in large area. In 3D painting, this problem manifests itself at much smaller brush size since a 3D brush can paint a larger number of voxels than the number of pixels in 2D painting. A solution is to use adaptive grid, where we can refine the grid only up to a resolution sufficient to represent the brush detail; for a smaller brush, we refine down to deep, but for a larger brush, we stop at a resolution that is roughly 10 voxels corresponding to the brush radius. When a brush is applied to an already refined region, this can still be slow. In this case, we coarsen the grid at the center of the brush where color should be merged into uniform brush color. Not to destroy exiting details, we coarsen rather conservatively than aggressively. This way, although painting is not always immediately responsive, but is always interactive and tolerable, even when artist paint a very large region in one stroke. The sky in Fig. \ref{fig:teaser} was painted with a very large brush (about a kilometer) whose effective radius in finest resolution is millions of voxels.

\subsection{Deferred Refining and Coarsening in Octree}
Although octree can effectively reject cells that do not intersect with brush stamps, painting a deep octree can be still expensive, when brush stroke is applied near highly-refined regions. We ease this problem by developing a multi-step strategy. We first paint on an existing tree without tree adjustment and update it on GPU. Most time, this strategy provides immediate visual feedback to user. The next step is a tree-adjustment stage. We mark cells that should be refined or coarsened and perform one -level refinement or coarsening per each frame. 
To update GPU textures, we split a texture into multiple sub-blocks, mark blocks dirty when CPU modifies them, and then update only the dirty blocks.

\subsection{Painting Interface}
Our volume painting system supports conventional painting tools for artists - such as color mixing, eraser, recoloring and color-blending. As a choice for painting user interfaces, we use off-the-shelf VR controllers such as HTC Vive controllers. When the controller is triggered, we record the location of controller and sub-sample the location at lower frequency (about 5Hz) to paint. 

\begin{figure}[htb]
    \centering
    \includegraphics[height=4.2cm]{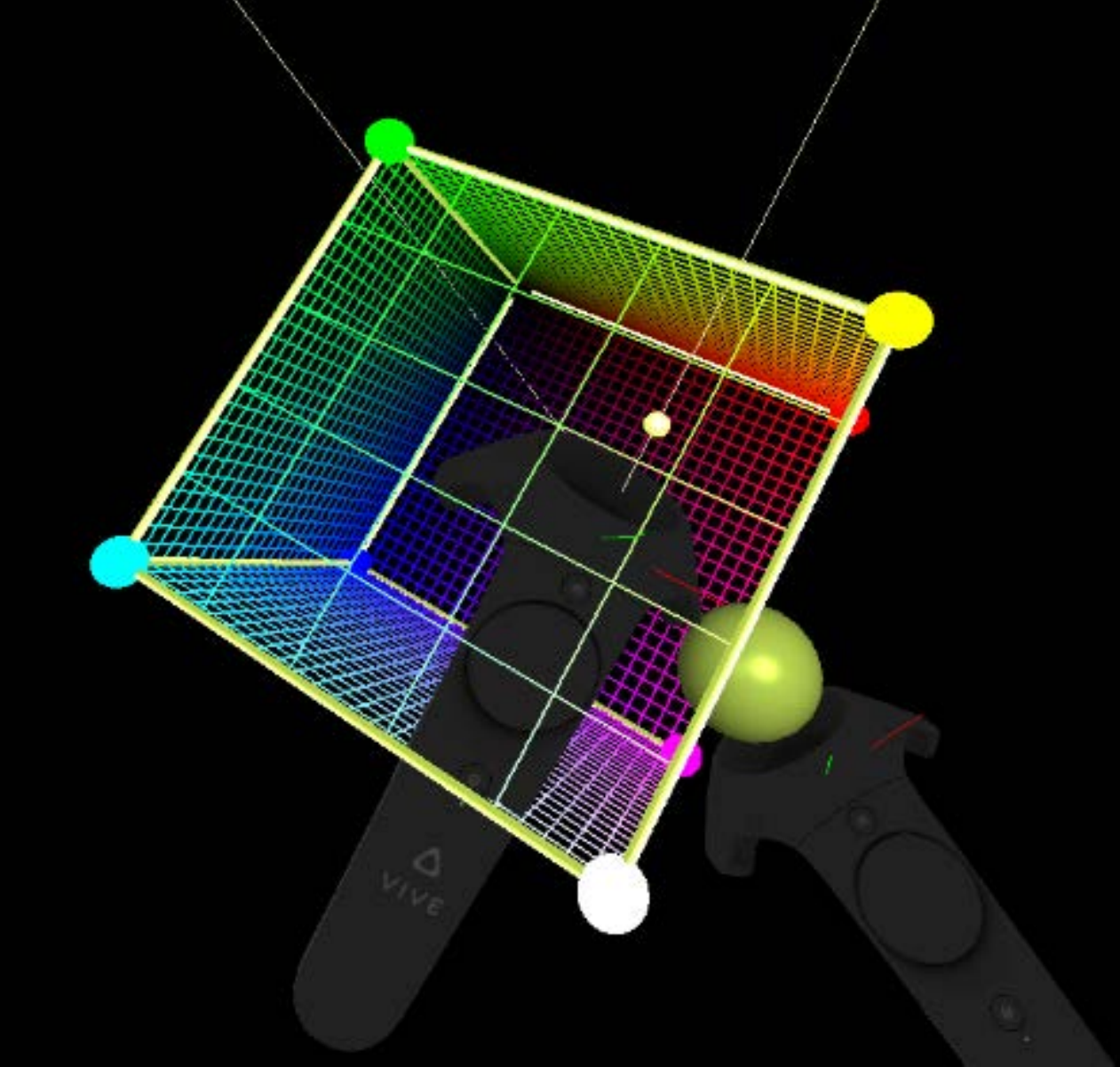}
    \includegraphics[height=4.2cm]{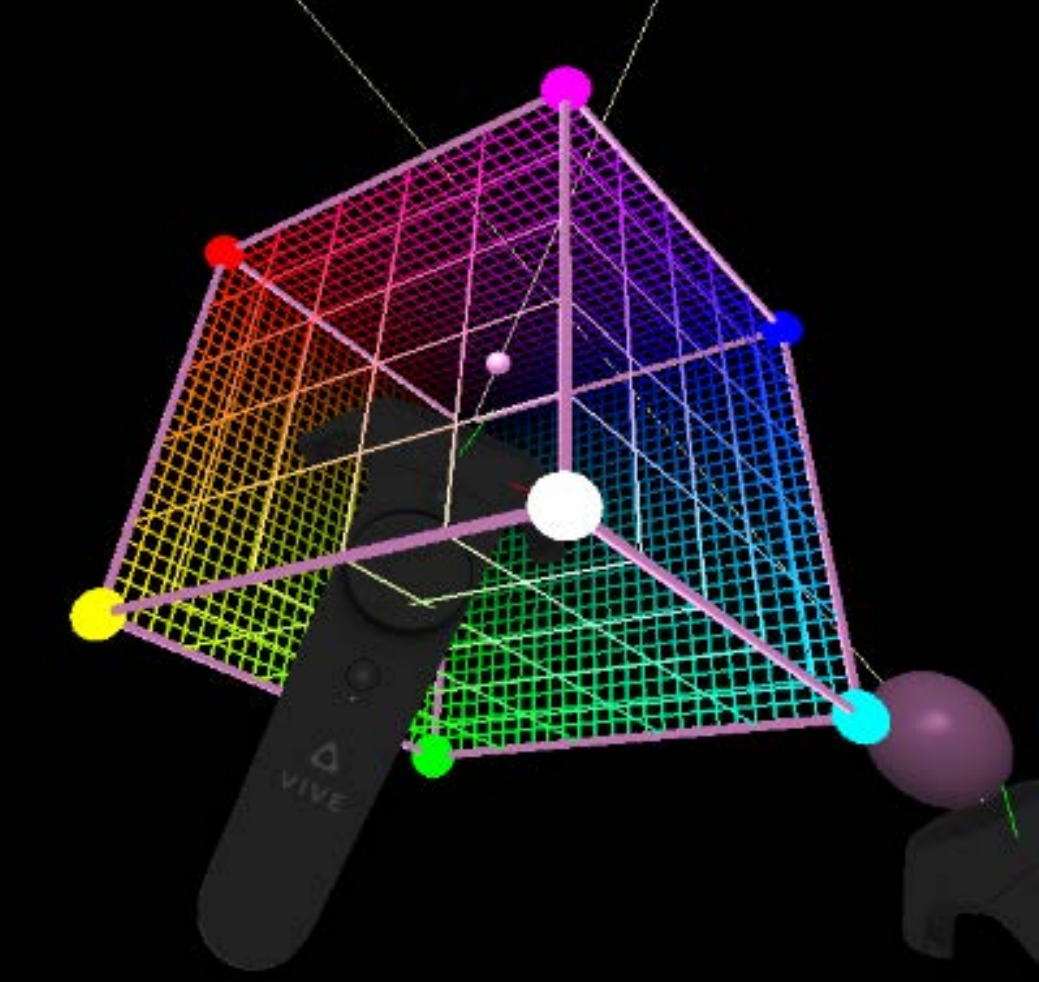}
    \includegraphics[height=4.2cm]{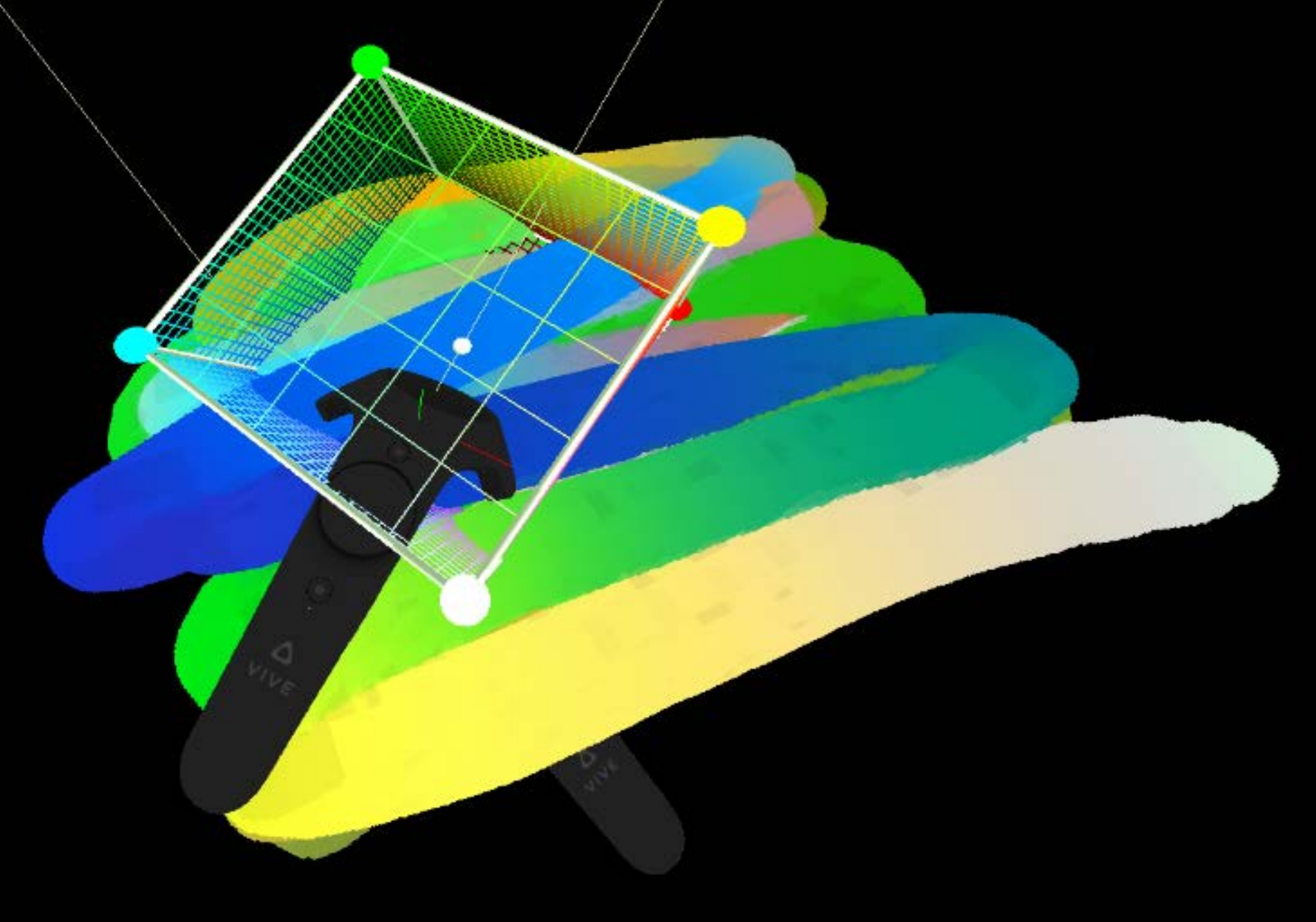}
\caption{One-handed color picker interface. Using only one hand, a user picks yellowish color (left), rotates the cube (middle), and changes color while applying strokes at the same time (right).}
\vspace{-0.5cm}
    \label{fig:colorui}
\end{figure}

Right-handed horizontal touch-swipe changes the painting transparency and the brush radius. The brush radius is scaled by the zoom level. So, to have a larger brush, the user needs to just zoom out. Trigger pressure also affects the brush radius.
As shown in Fig. \ref{fig:colorui}, we implemented a novel 3D color picking interface that allows user to select brush-color using a left-handed controller while drawing. Note that existing color pickers often require a user to use both of their hands, and do not allow changing color during stroking. Our color picker displays a RGB-color cube, and the user can move the controller inside the cube to change the brush color. The user can also rotate or translate the cube. For navigating canvas, painters use both hand controllers in a way that they grab some points in the canvas and move them around.

\section{Rendering Large Volumetric Canvas}
Another challenge lurks in rendering 3D volumetric paintings. One viable solution is to extract voxel faces and render them through raster graphics pipeline using, for instance, OpenGL similarly to Minecraft. However, as the number of grids in non-uniform size increases, the extracted vertex positions, particularly far away from the origin, may not be accurate due to numerical error. 
More significantly, geometric extraction requires a substantial amount of computational time, as the number of voxels grows. Consequently, we explore an alternative approach: ray casting the volumetric data in octree.

\subsection{Accurate Ray Starting Point}
To edit fine detail, artists should be able to zoom in to observe cells with highest depth (e.g. 24). However, the size of these tree cells can be even smaller than the single-precision floating point granularity except near the origin. Naively using floating point for  eye position in canvas coordinate will make head positions snapped to nearby floating point values, and more significantly, the eye distance will be erratic. This leads to extreme discomfort. Therefore, we carefully maintain canvas-to-VR, VR-to-HMD, and HMD-to-eye coordinate transformations so that the eye position precision is not lost. 

We propose to compute ray starting point in a cell local coordinate frame, and
keep the positioning error of the starting point sufficiently small because of the following reasons. Our cell-local coordinate system has origin at the cell center and has size one. The coordinates inside the cell has the range of [-0.5, 0.5]. Consequently, the finest resolution inside a cell is $0.5\times2^{-23}$ in single-precision floating point regardless of the size of the cell. If a ray starts from a leaf cell of depth 24, its resolution inside the leaf is extremely high. In contrast, if a 10km-sized root is not refined hence is a leaf, then the closest point to the boundary of the cell is 0.3mm away, merely equal to the width of a cell with depth 24. This appears to be coarse. However, users will not need to zoom in deep inside such a large cell. Even if a user zooms in, nearby cells will have depths at most one, not providing any detail that can be observed. Moreover, when a user starts painting, the cell will be refined, and ray starting point will be represented in a coordinate frame of the refined cell.

\begin{figure}
    \centering
    \includegraphics[width=0.69\linewidth]{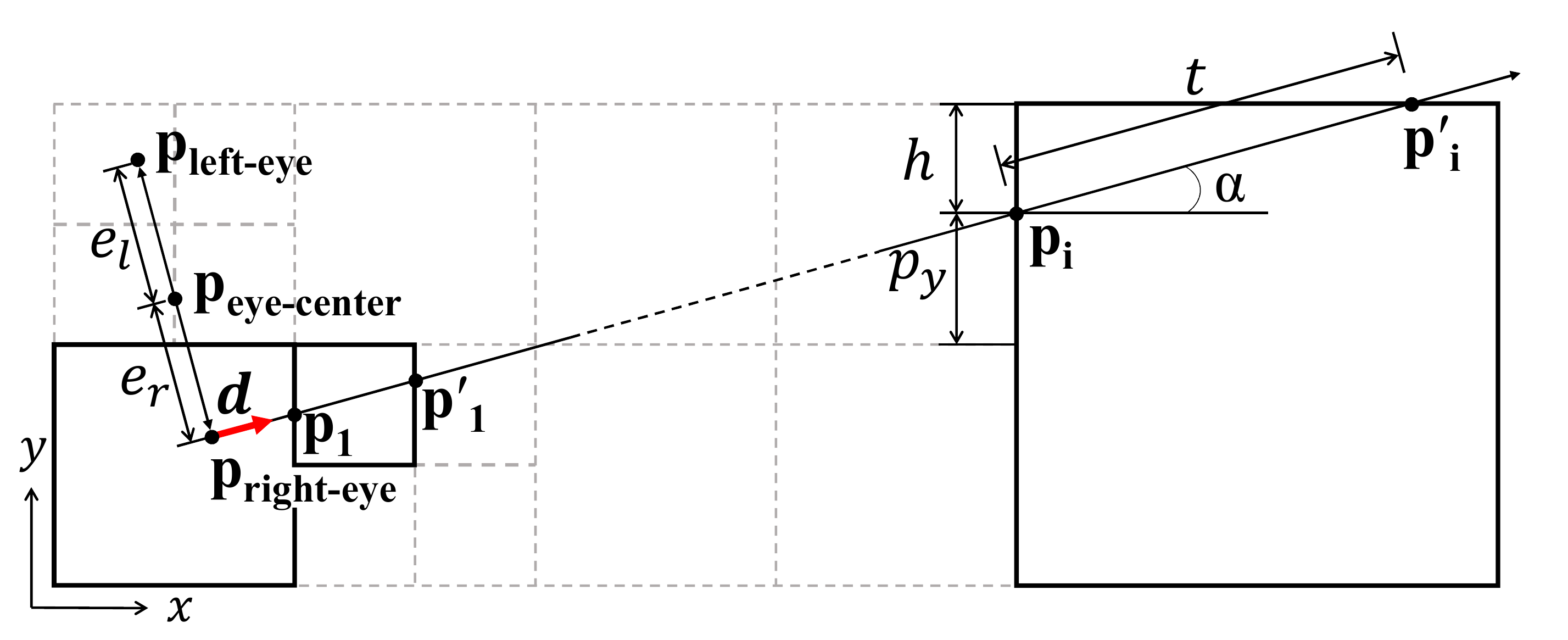}
    \caption{2D illustration of ray casting on adaptive grids. A ray is fired from the right eye of the user at $\mathbf{p}_\text{right-eye}$ along $\mathbf{d}$, and the first cell-entry and exit points $\mathbf{p}_0, \mathbf{p}\prime_0$ are calculated for the ray. This process continues for other cells until the ray is terminated.}
    \label{fig:rayaccuracy}
\end{figure}

\subsection{Accurate Ray Traversal}
As illustrated in Fig. \ref{fig:rayaccuracy}, given a ray and its direction $\mathbf{d}$, and a cell-entry point $\mathbf{p}_i$ of the ray into the $i^\text{\scriptsize th}$ cell, we compute the cell-traversal distance $t_i$ and the cell-exit point $\mathbf{p}_i\prime$ as well as a neighboring cell $N$ containing $\mathbf{p}_i\prime$. Since our volumetric canvas covers a large space and cell sizes vary by a large magnitude, the use of global coordinate system to calculate $\mathbf{p}_i\prime$ and $t$ can be highly inaccurate. In contrast, cell-local coordinate system can produce accurate results regardless of zoom level. 

Thus, we represent the cell-entry point $\mathbf{p_i}$ in terms of the cell-local coordinate system, where the origin is located at the cell center and the cell size is normalized to one. We also find the face that contains the cell-exit point $\mathbf{p\prime}_i$ with respect to the cell coordinate system. Using the intersecting face and the neighbor texture described in section \ref{sec:treetopologies}, we  choose the neighbor cell, the $i+1^\text{\scriptsize th}$ cell, to visit, and update $\mathbf{p}_i\prime$ to $\mathbf{p}_{i+1}\prime$. This process is repeated until the ray terminates after accumulating full opacity or exits the canvas.

\subsection{Analysis}\label{sec:rayErrorAnalysis}
As illustrated in Fig. \ref{fig:rayaccuracy}, in the $i^\text{\scriptsize th}$ cell, if the ray hits the top surface, the ray traversal-length $t_i$ is computed as $t_i = (0.5-p_y) / d_y$, where $p_y$ is the $y$ coordinate of $\mathbf{p}_i$, and $d_y$ is the $y$ component of $\mathbf{d}$. The error in $t_i$ will be proportional to $t_i$. Let the machine epsilon $\epsilon = 2^{-23}$ for a single floating point, and $f(x)$ be a floating point representation of $x$. Then $f(x+y) = (x+y)(1+\epsilon_+)$, with $\epsilon_+\le\epsilon$. Similarly, the error in $t_i$ is computed as
\begin{equation}
\begin{split}
    f(t_i) &=f\left( \frac{f(0.5-p_y)} {d_y}\right) = \frac{(0.5-p_y)(1+\epsilon_1)} {d_y}(1+\epsilon_2) \\
    &= t_i(1+\epsilon_1+\epsilon_1 +\epsilon_1\epsilon_2)= t_i (1+2\epsilon_t),   \;\;\;\epsilon_1, \epsilon_2 \le \epsilon.
\end{split}
\end{equation}
Note that ignoring $\epsilon_1\epsilon_2$, we have $\epsilon_t\le\epsilon$. We then compute $f(\mathbf{p}_i\prime) = f(\mathbf{p}_i + f(t_i)) = f(\mathbf{p}_i + t(1+2\epsilon_t)) = (\mathbf{p}_i + t_i(1+2\epsilon_t))(1 + \epsilon_3)) = (\mathbf{p}_i + t_i)(1+3\epsilon_p)$, for some $\epsilon_p \le\epsilon$, ignoring $\epsilon_3\epsilon_t$. Next, we transform the coordinates of $f(\mathbf{p}_i\prime)$ to the neighboring $i+1^\text{\scriptsize th}$ cell where the ray continues. This point is computed as $\mathbf{p}_{i+1} = f(f(f(\mathbf{p}_i\prime) s) + c) = ((\mathbf{p}_i + t_i)s+c)(1+5\epsilon_i)$ for some $\epsilon_i\le\epsilon$, where scale $s$ and shift $c$ depends on the depth and location. This way, $f(\mathbf{p}_{i+1}) = \tilde{\mathbf{p}}_{i+1}(1+5\epsilon)$, where $\tilde{\mathbf{p}}_{i+1}$ is the exact value computed from $\mathbf{p}_{i}$. Thus, the numerical error added during the traversal point is proportional to the coordinate values, the number of floating point operations $5$, and $\epsilon$.

Since we are using  cell coordinate system, each coordinate of $\mathbf{p}_i$ is in [-0.5,0.5]. Therefore, the error is always bounded by $2.5\epsilon$. In global coordinate system, the error is bounded by $2.5\epsilon w_i$, where $w_i$ is the size of the $i^{\text{\scriptsize th}}$ cell along the ray. Let $e_0$ be the error in eye location, i.e., the error introduced to compute $\mathbf{p}_\text{\scriptsize right-eye}$ in cell-local coordinate frame. Starting from this initial error $e_0$, traversing $n$ cells results in total error $e_0 + \sum_{i=0}^{n} 2.5 \epsilon w_i = e_0 + 2.5 \epsilon \sum_{i=0}^{n} w_i \le e_0 + 7.5 \epsilon L$, where $L$ is the ray length. Note that $\sum_{i=0}^{n} w_i \le 3L$. Thus, by using the cell coordinate system for ray/voxel traversal, we have shown that the error is proportional to $L$. Moreover, the error bound in angle $\sin^{-1}(\sqrt{2}(e_0/L + 7.5 \epsilon))$ does not increase as a function of $L$, and consequently the ray does not deviate from pixel center by more than a small fixed angle, regardless of the length of the ray $L$. 

\begin{figure}[htb]
    \centering
    \subfigure[Test Scene]{\includegraphics[width=0.25\linewidth]{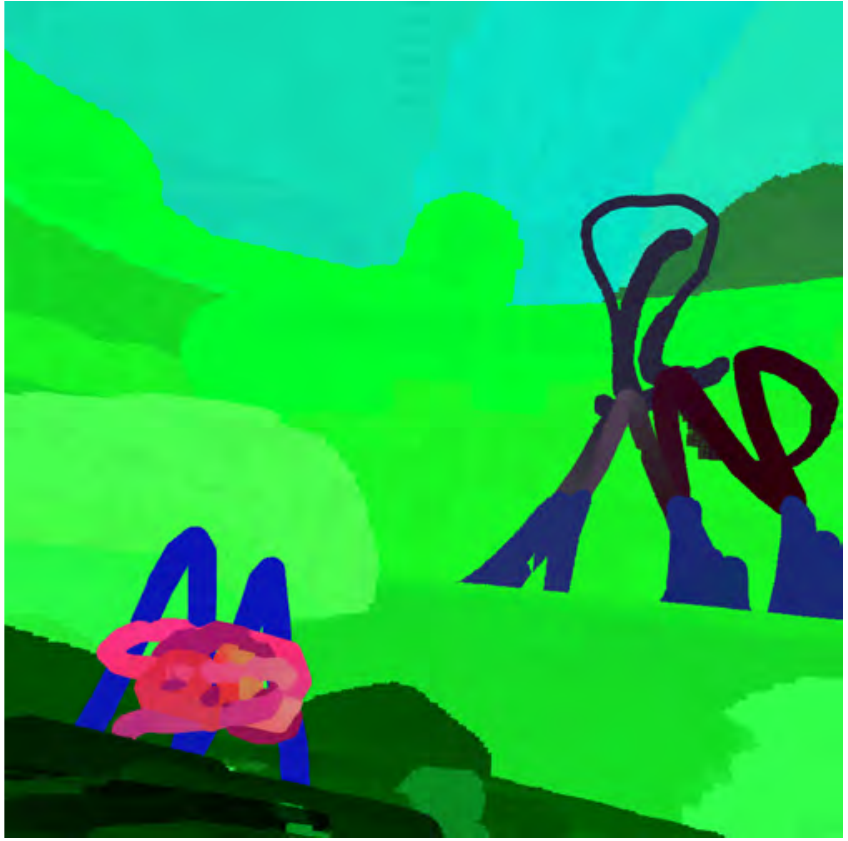}}%
    \subfigure[Errors in Ray Angle]{\includegraphics[width=0.25\linewidth]{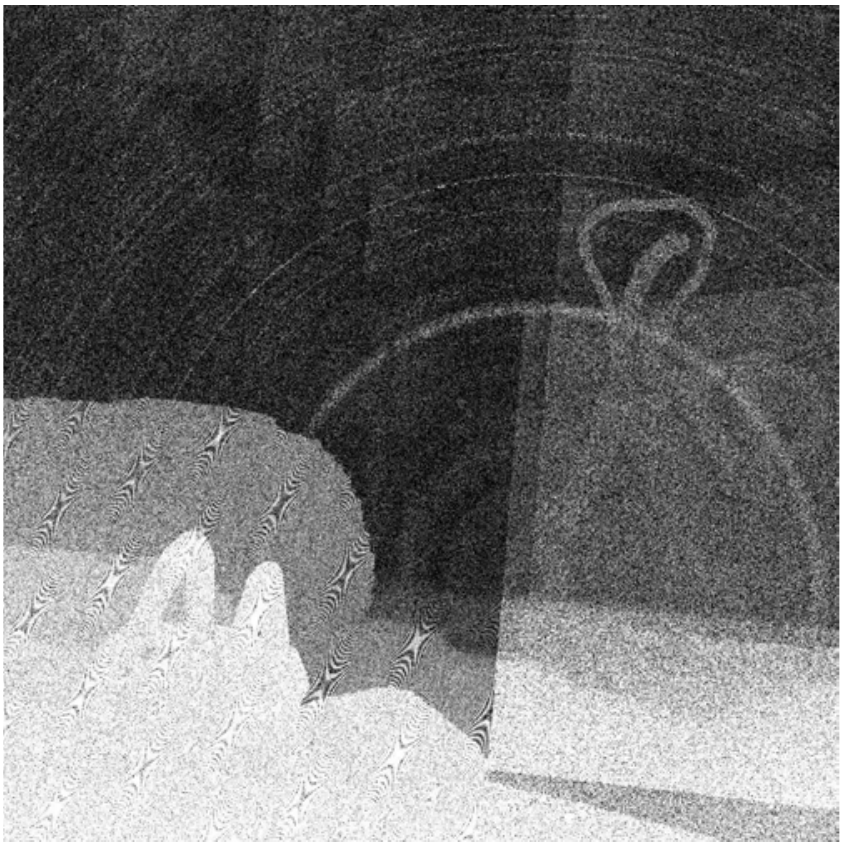}}%
    \subfigure[\# Of Crossed Cells]{\includegraphics[width=0.25\linewidth]{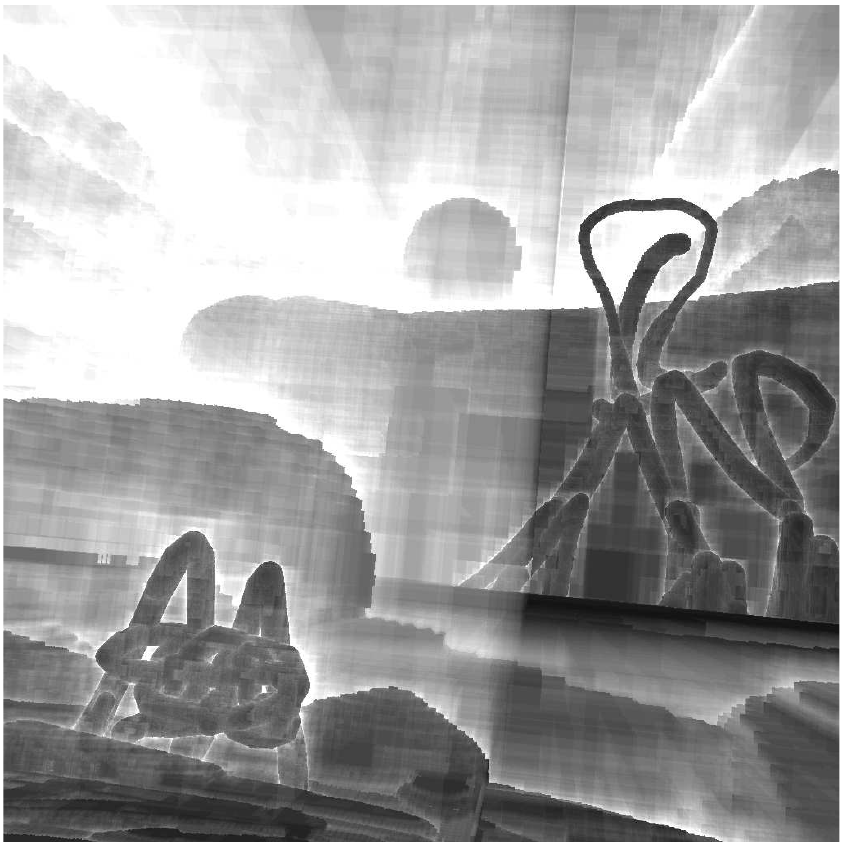}} 
    \subfigure[Ray Length VS Angle Error]{\includegraphics[width=0.375\linewidth]{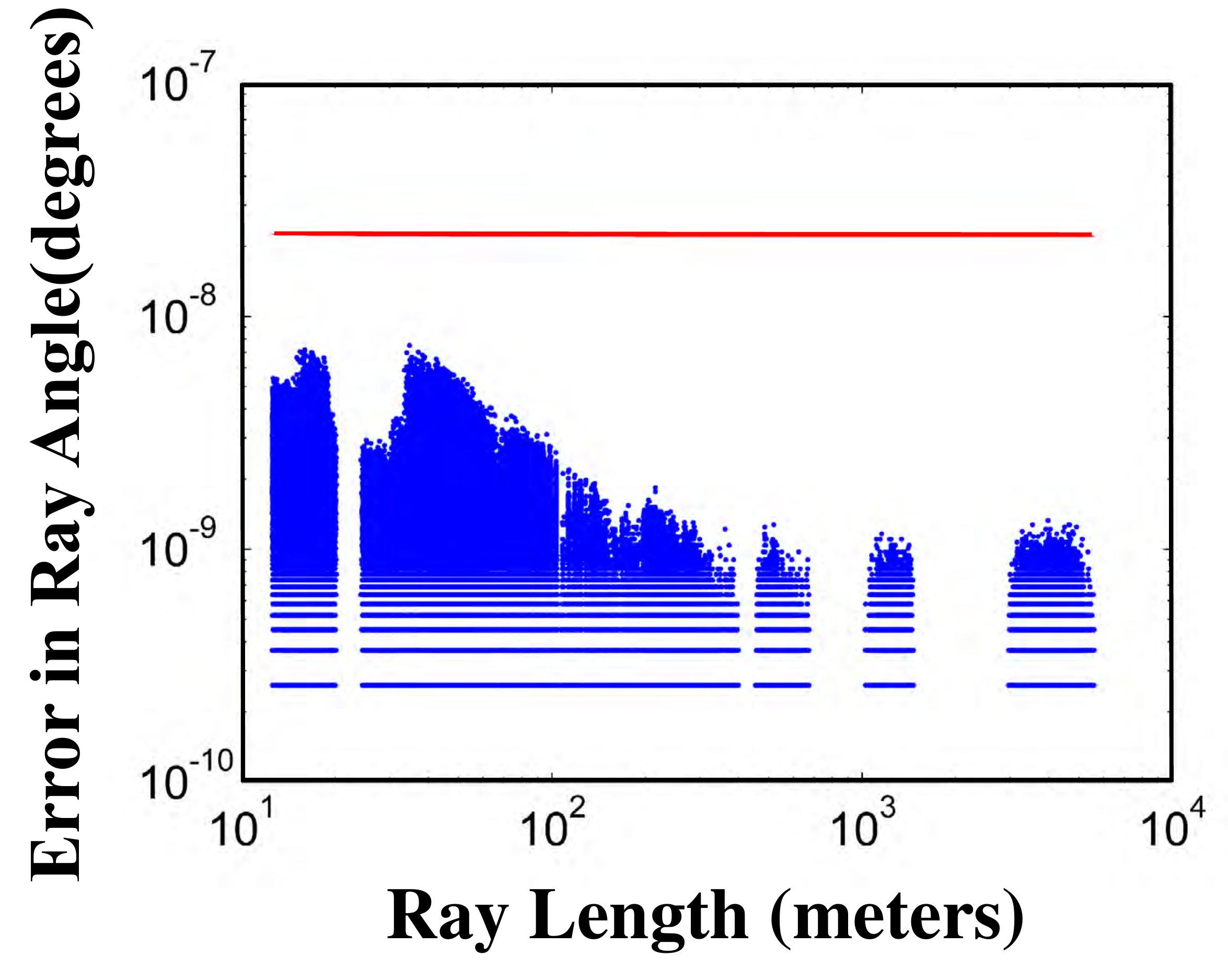}}
    \subfigure[Crossed Cells VS Angle Error]{\includegraphics[width=0.375\linewidth]{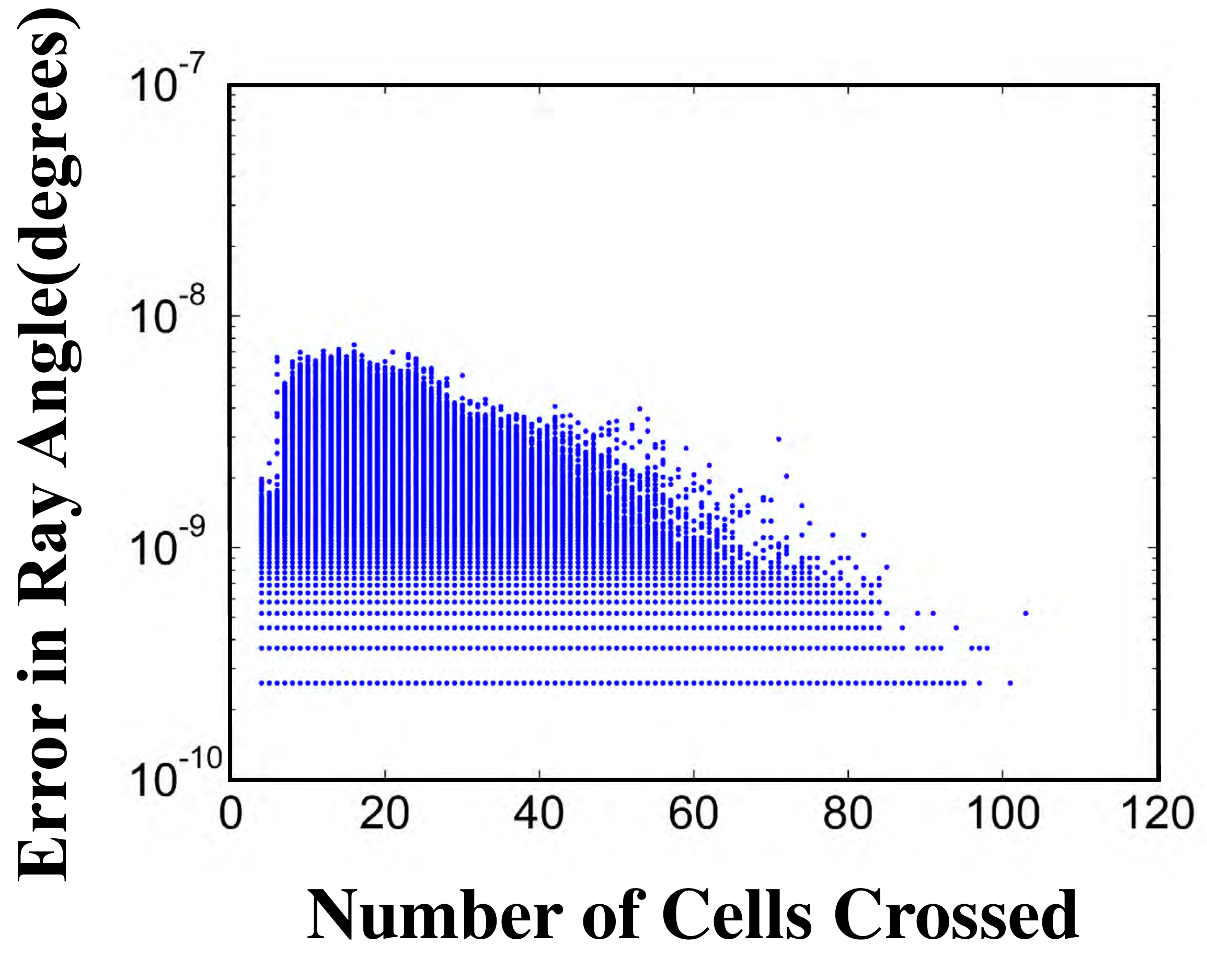}}      
    \caption{Ray traversal using local coordinate system. 
    (a) Test VR painting, (b) Error angle between the final ray location and the initial ray direction, magnified by $10^9$ for display in gray. (c) Number of cells crossed in gray. (d) and (e) show that the ray-angle error does not grow as the ray traverses. In (d), the initial error $e_0/L$ is dominant when $L$ is small. In (e), the more cells the ray crosses, we see the more stochastic decay. The red line in (d) is the worst case error bound. 
    }
    \label{fig:rayAngleError}
\end{figure}

To verify, we performed an experiment, as demonstrated in Fig. \ref{fig:rayAngleError}, and show that the screen-space ray-deviation from pixel center, formulated as ray-angle error, does not accumulate during ray traversal. In fact, the error is indeed very small, and is relatively larger in nearby pixels (the maximum ray-angle error is $7.5\times10^{-9}$ degrees) due to the initial position error (the position error is $2.5\times10^{-7}$ in $L_2$ norm) and then slightly decays as $L$ increases. This experimental result implies that we can perform ray-casting even on mobile GPUs with only half-precision floats ($\epsilon=1/1024$) using fragment shaders.

\begin{figure}[htb]
    \centering
    \subfigure[Canvas Coordinate]{\includegraphics[height=4.5cm]{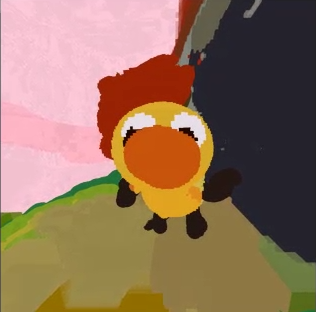}}  \hspace{0.5cm} 
    \subfigure[Cell Coordinate]{\includegraphics[height=4.5cm]{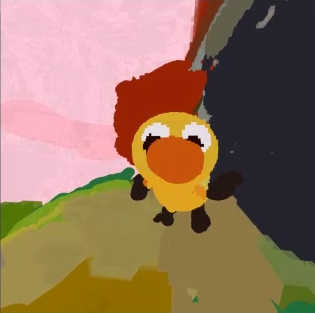}}     
    \caption{When canvas (world) coordinates were used, a ray drifts while traversing cells. This results in a large angle error (a). We show that this problem is completely solved by using cell-local coordinates (b). }
    \label{fig:accuracycomparison}
    \vspace{-0.5cm}
\end{figure}

We compare the results using world coordinate and local coordinate in Fig. \ref{fig:accuracycomparison}.

\section{Results and Discussions}

We have implemented our prototype VR painting system, {\em CanvoX}, that utilizes deep octrees to represent large volumetric canvas and accurate ray-casting algorithms for volume rendering. Our hardware setup includes Nvidia GTX 980Ti GPU and Intel Core i7-4790 CPU with 16GB RAM for rendering computation and HTC Vive for interfacing immersive and personal VR environments. 

\subsubsection*{Volume Painting Results}

\begin{figure*}[htb]
    \centering
    \includegraphics[height=4.cm]{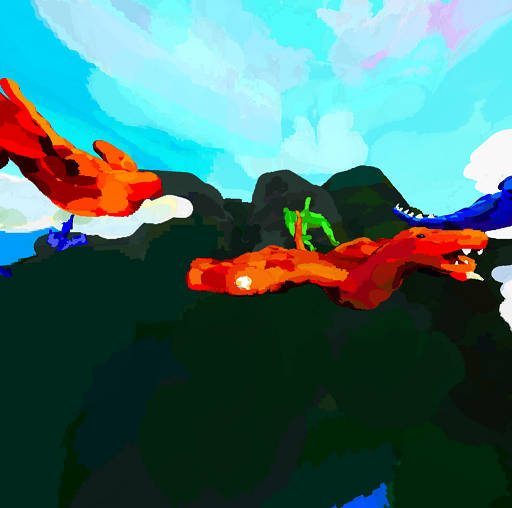}
    \includegraphics[height=4.cm]{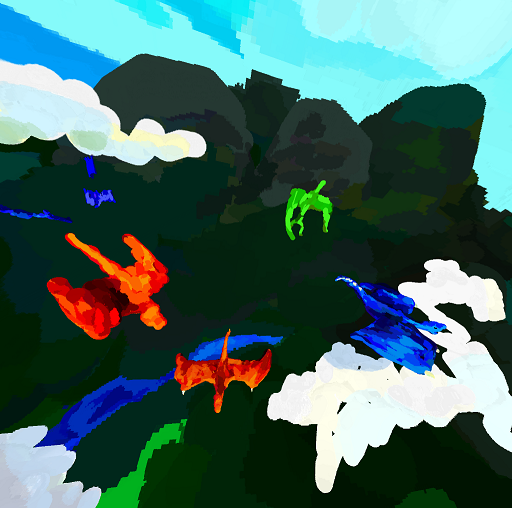}
    \includegraphics[height=4.cm]{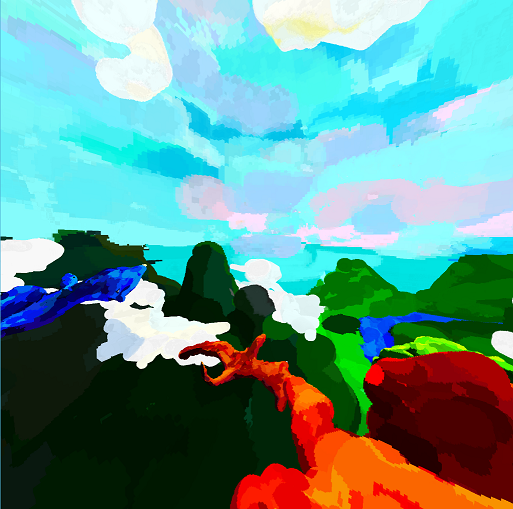}
    \includegraphics[height=4.cm]{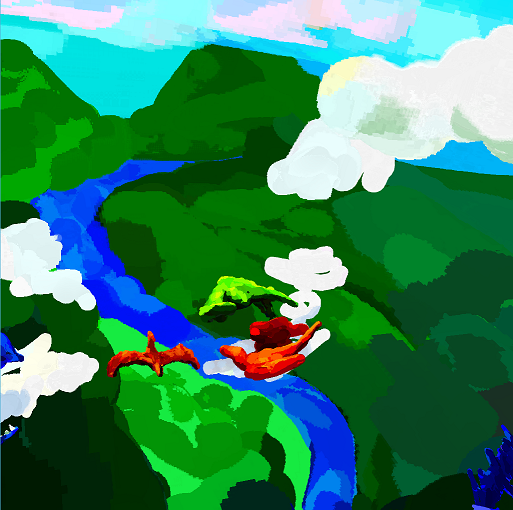}\\
    \includegraphics[height=4.cm]{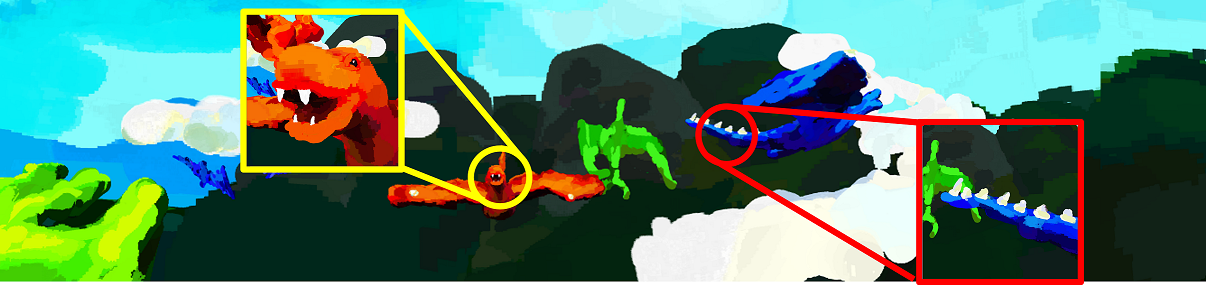}
    \caption{"Flying dragons" in different view points (top), from a single 3D painting (bottom)}
\label{fig:dragons}
\end{figure*}

\begin{figure*}[htb]
    \centering
    \subfigure{\includegraphics[height=4.0cm]{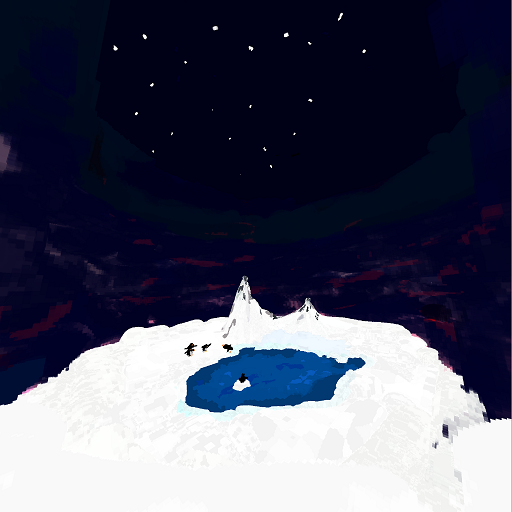}}
    \subfigure{\includegraphics[height=4.0cm]{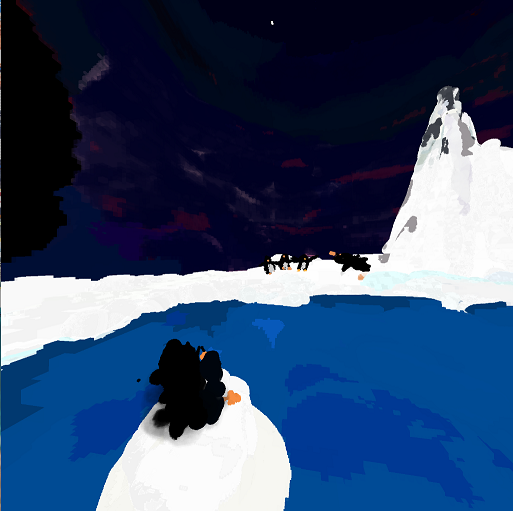}}
    \subfigure{\includegraphics[height=4.0cm]{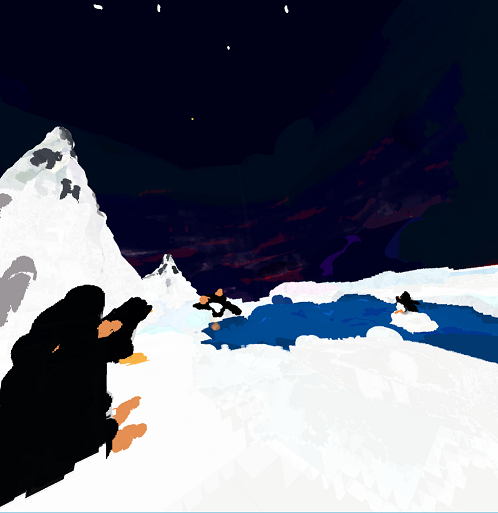}}
    \subfigure{\includegraphics[height=4.0cm]{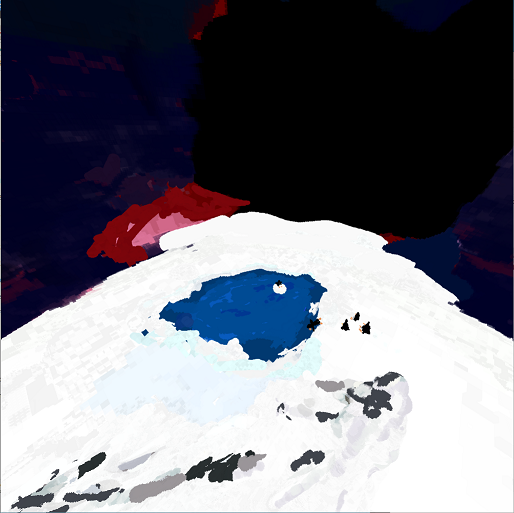}}
\caption{"Snow Mountain" in different view points.}
\label{fig:snowmountain}
\end{figure*}

\begin{figure*}[htb]
    \centering
    \subfigure{\includegraphics[height=4.0cm]{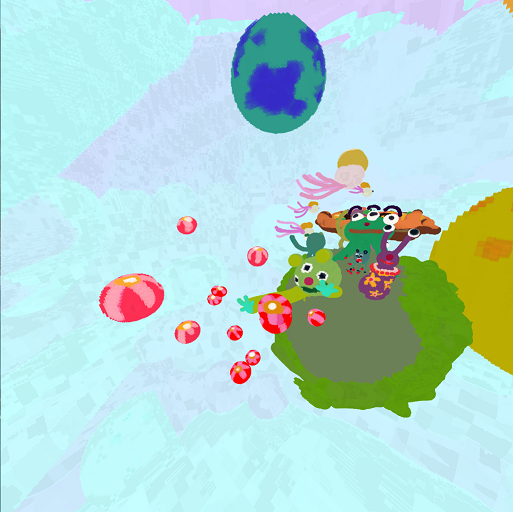}}
    \subfigure{\includegraphics[height=4.0cm]{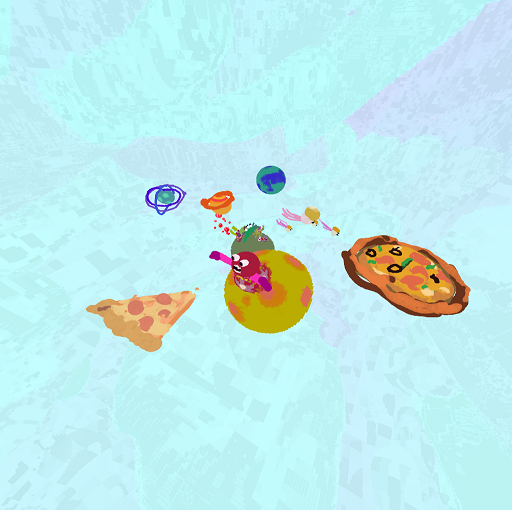}}
    \subfigure{\includegraphics[height=4.0cm]{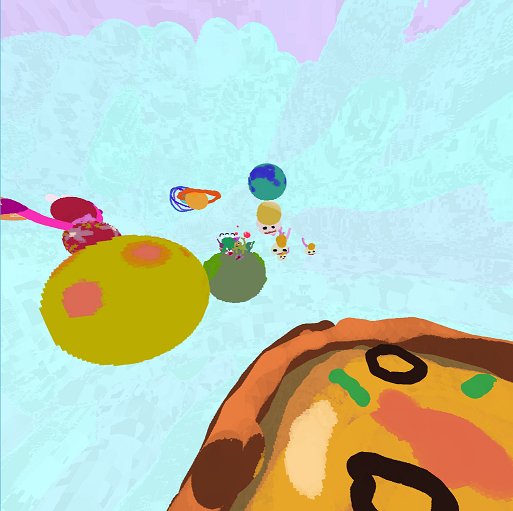}}
    \subfigure{\includegraphics[height=4.0cm]{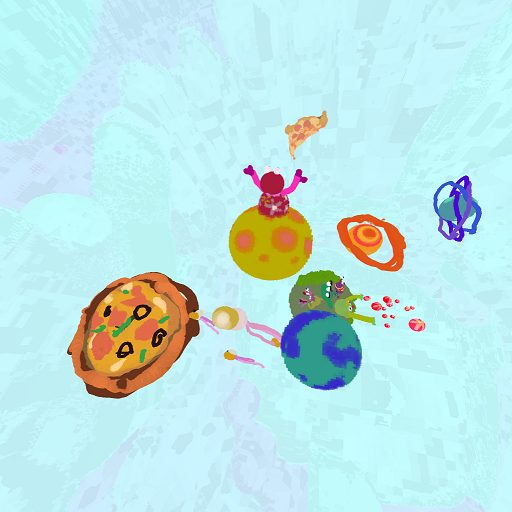}}
\caption{"Imaginary Island" in different view points.}
\label{fig:imaginaryworld}
\end{figure*}

We invited digital-painting artists to produce five pieces of volumetric paintings. 
"Floating Island and B-612" in Fig.~\ref{fig:teaser} represents a good example as extension of 2D painting to 3D. The artist painted the scene with two distinct appreciation rooms with large and colorful sky background. Inside the room, objects such as little duck, flying birds and apples also show the benefits of our system. By adding rough shades using recolor mode, the artist drew a fairy-tale island which can be appreciated from any view point. In a similar way, "Snow Mountain"(Fig.~\ref{fig:snowmountain}) also shows that an artist can easily draw the mountain in background and paint detailed objects such as penguins in the room. 
"Flying dragons"(Fig.~\ref{fig:dragons}) used more rough and colorful shades on dragon's body-surface but still maintained fine details on eyes, teeth and horns. This result show that our system allows artists to paint a large-scale 3D scene to their full control. Moreover, they can set multiple interesting appreciation rooms at, for instance, back on each dragon, top of the mountain, or inside clouds. The recoloring and color-mix tools were used for painting naturally-blended sky. In "Island"(Fig.~\ref{fig:teaser}) and "Imaginary World"(Fig.~\ref{fig:imaginaryworld}), artist explores much larger space to paint outside the room. 

Interesting remarks from artists were that they have to change the way they used to perceive painting and that they need to step away from the familiarity of perspectives on 2D canvas. The second remark is very interesting as it appears to be result of being able to paint in very large 3D canvas. We believe future explorations with artist will reveal how we can paint perspective in 3D canvas. For example, recoloring far away mountains from foreground room would be an interesting way of painting to explore.


\section{Conclusions}

In this paper, we have proposed a new type of 3D art, {\em volumetric VR painting},  that allows artists to express artistic details on a large volumetric canvas in VR environment. To represent the large volumetric canvas, we adopt a CPU/GPU hybrid representation of deep octrees with a wide range of varying resolution, and also exploit the hybrid representation to share the painting workloads so that the system can run at interactive rates. 
Moreover, we address the problem of ray-casting precision for deep octrees and analyze it.
Our 3D volume painting system is just the beginning of a new art form, and thus we have a few issues/limitations that we would like to address in future. First of all, our system still consumes memory significantly due to the sheer size of volumetric data. We plan to port our system to mobile VR platforms such as Google daydream or GearVR using our accurate ray casting techniques. 
We will also need to improve our painting interfaces to be more versatile, competitive to existing 2D digital painting systems. Finally, we plan to release our volume painting system to the public to initiate and lead a new community for  volume painting.


\bibliographystyle{acmsiggraph}
\bibliography{vpainting}
\begin{figure*}[htb]
    \centering
    \subfigure{\includegraphics[height=4.0cm]{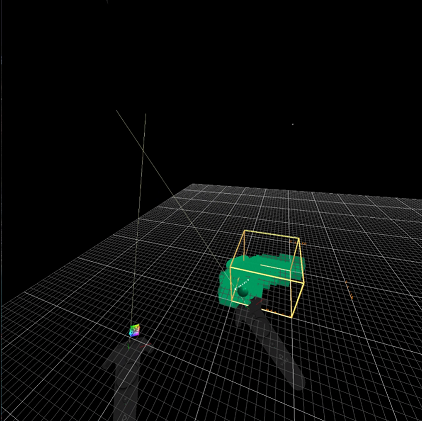}}%
    \subfigure{\includegraphics[height=4.0cm]{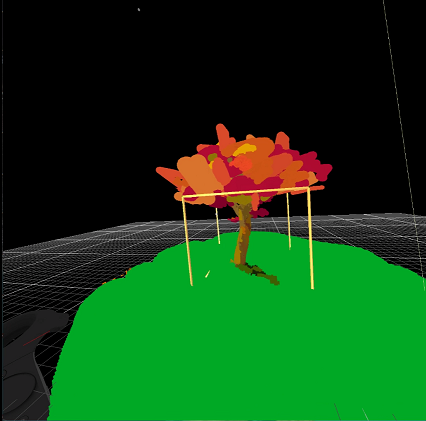}}
    \subfigure{\includegraphics[height=4.0cm]{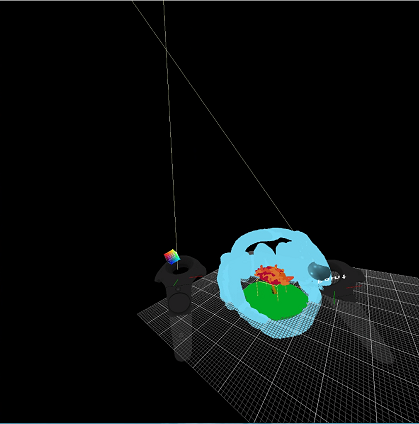}}%
    \subfigure{\includegraphics[height=4.0cm]{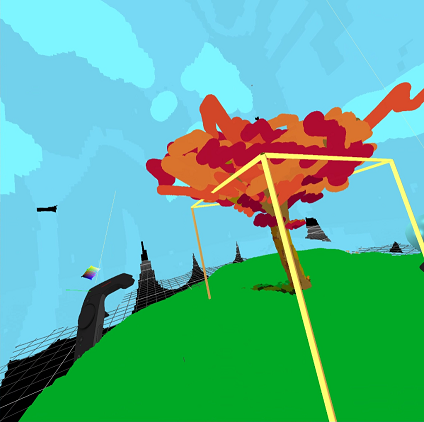}}
    \caption{A typical painting procedure. (a) painting ground, (b) painting foreground object, (c) painting background, (d) painting continues inside the background.}
    \label{fig:drawingTutorial}
\end{figure*}

\end{document}